\documentclass[10pt, preprint]{aastex63}
\usepackage{graphicx,amsmath,amssymb,bm,tabularx}
\usepackage{enumerate}
\usepackage{isotope}
\usepackage{multirow,dcolumn,booktabs}
\usepackage{bbm}
\usepackage{braket}

\allowdisplaybreaks[1]

\newcommand{\fmiq}{\, \text{fm}^{-3}}

\newcommand{\mev}{\, \text{MeV}}

\renewcommand{\d}{\mathrm{d}}

\newcommand{\km}{\,\mathrm{km}}

\newcommand{\beq}{\begin{equation}}
\newcommand{\eeq}{\end{equation}}

\allowdisplaybreaks

\shortauthors{Tews and Schwenk}

\begin{document}

\title{Spin-polarized neutron matter, the maximum mass of neutron stars, and GW170817}


\correspondingauthor{I.\ Tews}
\email{itews@lanl.gov}

\author{I.\ Tews}
\affiliation{Theoretical Division, Los Alamos National Laboratory, Los Alamos, NM 87545, USA}

\author{A.\ Schwenk}
\affiliation{Institut f\"ur Kernphysik, Technische Universit\"at Darmstadt, 64289 Darmstadt, Germany}
\affiliation{ExtreMe Matter Institute EMMI, GSI Helmholtzzentrum f\"ur Schwerionenforschung GmbH, 64291 Darmstadt, Germany}
\affiliation{Max-Planck-Institut f\"ur Kernphysik, Saupfercheckweg 1, 69117 Heidelberg, Germany}

\begin{abstract}
We investigate how a phase transition from neutron-star matter to spin-polarized neutron matter affects the equation of state and mass-radius relation of neutron stars. While general extension schemes for the equation of state allow for high pressures inside neutron stars, we find that a phase transition to spin-polarized neutron matter excludes extreme regimes. 
Hence, such a transition limits the maximum mass of neutron stars to lie below 2.6$-$$2.9 \, M_{\odot}$, depending on the microscopic nuclear forces used, while significantly larger masses could be reached without these constraints. 
These limits are in good agreement with recent constraints extracted from the neutron-star merger GW170817 and its electromagnetic counterpart. Assuming the description in terms of spin-polarized neutron matter to be valid in the center of neutron stars, we find that stars with a large spin-polarized domain in their core are ruled out by GW170817.
\end{abstract}

\keywords{dense matter --- equation of state --- stars: neutron}

\section{Motivation}

Neutron-star observations, such as the recent detection of two merging neutron stars in the gravitational-wave (GW), gamma-ray, and electromagnetic (EM) spectra~\citep{Abbott:2017,GBM:2017lvd,Monitor:2017mdv,Abbott:2018wiz}, designated as GW170817, GRB 170817A, and AT 2017gfo, respectively, provide crucial constraints on the equation of state (EOS) of dense strongly interacting matter. The EOS is a key quantity for astrophysics and sensitively depends on strong interactions. Hence, it connects astrophysical observations to laboratory experiments at rare isotope beam facilities for the
most neutron-rich extremes, e.g., at the Radioactive Isotope Beam Factory (RIBF), Japan, and the future Facility for Rare Isotope Beams (FRIB) and Facility for Antiproton and Ion Research (FAIR) in the US and Germany.
While there is a wealth of theoretical models for the EOS of neutron-star matter (see \cite{Hebeler:2015hla,Lattimer:2015nhk,Ozel:2016oaf} for reviews), 
for densities beyond nuclear saturation density, $n_{\rm sat} \approx 0.16 \fmiq$, these models can only be confronted with a limited set of experimental data, e.g., from heavy-ion collisions~\citep{Danielewicz:2002pu}. 

Neutron stars are the densest objects in the universe and probe the EOS up to several times saturation density. Neutron-star observations are therefore an ideal source of additional information that complement experimental data and provide powerful constraints for the EOS at higher densities~\citep{Hebeler:2013nza,Tews:2018kmu}.
The structure of a neutron star is described by the mass-radius ($M$--$R$) relation, which is an important observational quantity and in a one-to-one correspondence with the EOS: the $M$--$R$ relation follows from the EOS by solving the Tolman-Oppenheimer-Volkoff equations~\citep{Tolman:1939jz,Oppenheimer:1939ne}. 
Measuring the $M$--$R$ relation, and therefore the EOS, observationally is however extremely difficult.

On the one hand, neutron-star masses can be determined very precisely for some neutron stars in binaries~\citep{Lattimer:2012nd}. For example, the precise measurement of two-solar-mass neutron stars~\citep{Demorest2010,Antoniadis2013,Fonseca2016} established a robust and strong constraint on the EOS of strongly interacting matter, and implied that the EOS has to be sufficiently stiff at high densities to support neutron stars in that mass regime. This constraint was recently tightened by the observation of a $2.14^{+0.10}_{-0.09}\,M_{\odot}$ neutron star~\citep{Cromartie:2019kug}. 
In contrast to masses, radii are extremely difficult to measure because of a limited number of suitable neutron stars and large systematic and statistical uncertainties~\citep{Steiner:2010fz}. Recently, the NICER~\citep{NICER1} mission was able to measure simultaneously the mass and radius of PSR J0030+0451, with a radius $12.7^{+1.1}_{-1.2}$~km~\citep{Riley:2019yda} and $13.0^{+1.2}_{-1.1}$~km~\citep{Miller:2019cac} at the 68 \% confidence level. These results are consistent with our present understanding of the nuclear EOS \citep{Raaijmakers:2019qny}. Future observations by the NICER and, e.g., eXTP missions~\citep{Watts:2018iom} will improve this with target radius uncertainties of 5$-$10\,\%, corresponding to 1\,km or better~\citep{Watts:2016uzu}.

In this situation, the recent observation of a neutron-star merger~\citep{Abbott:2017,Abbott:2018wiz} by the LIGO-Virgo collaboration, as well as many many follow-up observations, have provided exciting additional insights. 
While the GW signal GW170817 has been used to constrain the radius of a typical $1.4 \,M_{\odot}$ neutron star to be below 13.6\,km (see also~\cite{Annala:2017llu,Most:2018hfd,Tews:2018iwm}), additional information can be obtained from the observed EM kilonova. 
Extracted ejecta properties disfavor a prompt collapse to a black hole, favoring a hypermassive neutron star supported by differential rotation as an immediate product of the merger. This object then collapsed to a black hole on the timescale of a few 100\,ms, because a longer-lived supramassive (supported against collapse by uniform rotation) or stable neutron star would have been able to deposit large amounts of rotational energy into the ejecta, leading to the formation of an energetic relativistic jet, which was not observed~\citep{Margalit:2017,Shibata:2017xdx,Rezzolla:2017aly}.

Based on the EM observation and the previously discussed scenario, one can propose limits on the maximum mass of non-rotating neutron stars,
$M_{\rm{max}}$. In general, a larger $M_{\rm{max}}$ leads to a larger
maximum mass for a rotating neutron star and, therefore, a longer lifetime of the merger remnant. The absence of a prompt collapse requires $M_{\rm{max}}$ to be sufficiently large to stabilize the hypermassive neutron star, while the absence of a longer-lived
remnant forces $M_{\rm{max}}$ to be sufficiently small~\citep{Margalit:2017}.
Based on this reasoning, several estimates on the upper limit of $M_{\rm{max}}$ were proposed. 
From the energy deposited into the kilonova ejecta, \cite{Margalit:2017} concluded 
$M_{\rm{max}}$ to be bounded by $M_{\text{max}} \leqslant 2.17 \, M_{\odot}$, similar to \cite{Shibata:2017xdx} who found $M_{\text{max}}=2.2 \pm 0.05 \, M_{\odot}$. This constraint was recently updated to $M_{\text{max}} \leqslant 2.3 \, M_{\odot}$ \citep{Shibata:2019ctb}. \cite{Rezzolla:2017aly} used empirical relations between $M_{\rm{max}}$ and the maximum masses of uniformly rotating or differentially rotating neutron stars to conclude $M_{\text{max}} \leqslant 2.16^{+0.17}_{-0.15} \, M_{\odot}$, and \cite{Ruiz:2017due} found $M_{\text{max}}=2.16$$-$$2.28 \, M_{\odot}$. 

From the theoretical side, the range of predicted $M_{\rm{max}}$ varies over a much wider range. Model EOSs for astrophysical simulations typically have
$M_{\rm{max}}\sim 2-2.5\,M_{\odot}$~\citep{Lattimer:2012nd}, 
while EOSs based on modern nuclear forces at nuclear densities and general extrapolations 
to higher densities usually allow for a much wider range of pressure in neutron stars~\citep{Hebeler:2013nza,Kruger:2013kua,Annala:2017llu,Tews:2018iwm,Greif:2018njt}, and can support
extreme values for $M_{\rm{max}}$, limited only by $M_{\rm{max}} \lesssim 4.0 \, M_\odot$ ($M_{\rm{max}}\lesssim 2.9 \, M_\odot$) when a nucleonic EOS is considered up to $n_{\rm{sat}}$ ($2n_{\rm{sat}}$) in \cite{Tews:2018iwm}. 
Typically,
approaches using general extensions find $M_{\rm{max}}$ ranging between 2.9$-$$3.2\, M_\odot$~\citep{Hebeler:2013nza,Kruger:2013kua,Annala:2017llu,Greif:2018njt}, based on different sets of
calculations and other physically motivated constraints.
These ranges are consistent with earlier findings of $M_{\rm{max}}\lesssim 2.9 \, M_\odot$, where a nucleonic EOS up to $2n_{\rm{sat}}$ was combined with the stiffest possible EOS at higher densities~\citep{Nauenberg1973,Rhoades:1974fn}, or of $M_{\rm{max}}\lesssim 4.0 \, M_\odot$ in a similar approach, but considering a nucleonic EOS to a lower density~\citep{Kalogera:1996ci}.

\begin{figure*}[t]
\centering
\includegraphics[width=0.995\textwidth]{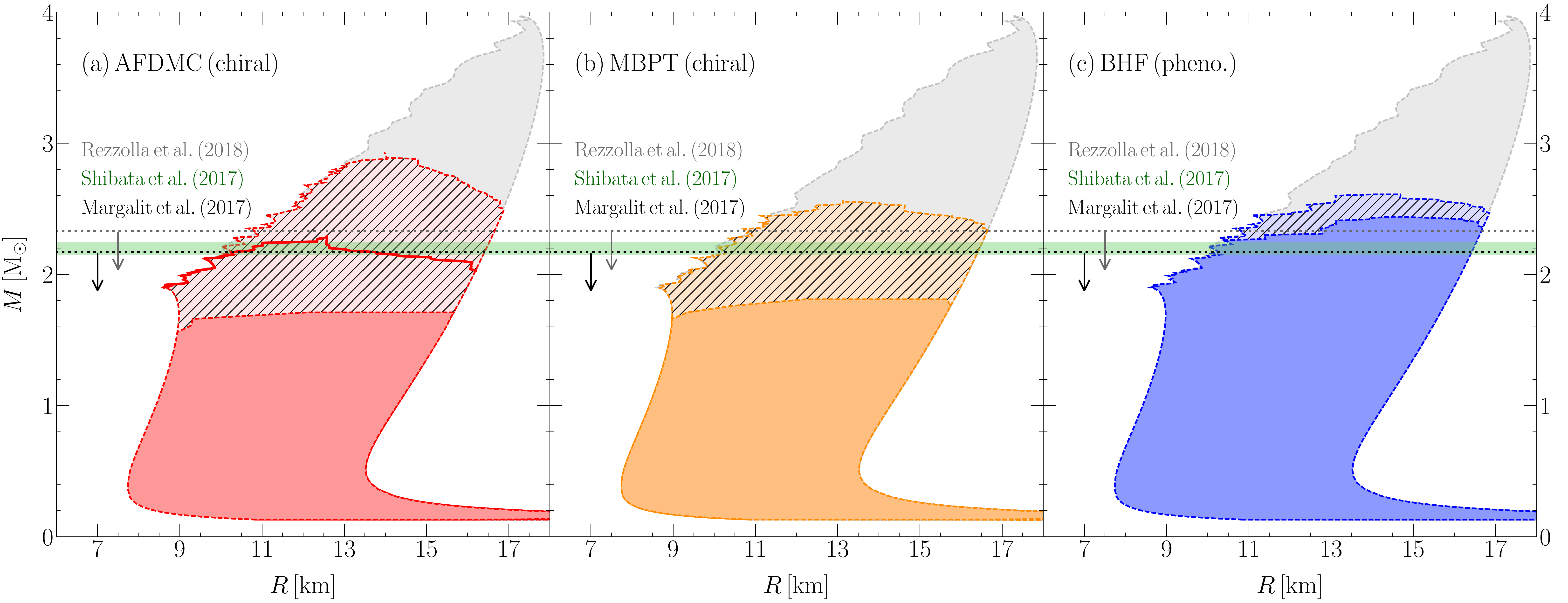}
\caption{\label{fig:MRplot_SpinPol}
Mass-radius relation for neutron stars using chiral EFT input up to $n_{\rm sat}$ and a speed-of-sound extension to higher densities (gray areas), and when considering a phase transition to SPM for (a) AFDMC calculations with local chiral interactions to N$^2$LO (red areas), (b) for MBPT calculations from \cite{Kruger:2014caa} based on chiral EFT interactions to N$^3$LO (orange areas), and (c) for BHF calculations based on the Nijmegen and Reid93 phenomenological interactions from \cite{Vidana:2001ei} (blue areas). The hatched areas correspond to the
uncertainty bands in the EOS of SPM,
and the solid red line in panel (a) to the centroid of the calculation.
The horizontal lines and band mark the inferred constraints from the EM signal of GW170817 by \cite{Margalit:2017}, \citep{Shibata:2017xdx}, and \citep{Rezzolla:2017aly}. 
}
\end{figure*} 

In this work, we propose a novel theoretical conjecture that limits $M_{\rm{max}} \leqslant 2.6-2.9 \, M_{\odot}$, and is therefore relevant for all extension schemes discussed above. 
In particular, we propose that a possible phase transition from unpolarized neutron-star matter to spin-polarized neutron matter (SPM) provides 
constraints on the properties of neutron-star matter and excludes areas with high pressure.
Our novel conjecture is only based on theoretical calculations of SPM, does not take any observational constraints on the maximum mass from GW170817 into account, and hence, is complementary to the observational conclusions.

We have investigated several EOSs for SPM and found that such a phase transition  
drastically softens the EOS, and therefore limits neutron-star masses.
This observation remains true for all possible unpolarized EOSs, even for those with regions of drastic stiffening beyond nuclear saturation density.
Only for the stiffest possible SPM EOS do we find stars that experience a spin-polarized phase in their core. 
However, in these cases, the resulting EOSs do not considerably increase the maximum neutron-star mass. Furthermore, most of the resulting EOSs are ruled out based on radius constraints from
GW170817~\citep{Abbott:2017,Annala:2017llu,Most:2018hfd,Tews:2018iwm,Abbott:2018wiz,Raaijmakers:2019dks}. 
Hence, we do not find any neutron stars consistent with astrophysical observations that exhibit a considerable domain of SPM in their core, in agreement with \cite{Vidana:2001ei}. 
Therefore, the onset of such a phase transition mainly determines the end of the stable branch of the neutron-star $M$$-$$R$ relation, limiting the range for $M_{\rm{max}}$.
We show our main findings in Fig.~\ref{fig:MRplot_SpinPol} for SPM calculated with various approaches that give very consistent results and limit $M_{\rm{max}}$ to be below 2.6$-$$2.9 \, M_{\odot}$. Our results exclude neutron stars that explore extreme pressures (gray areas in Fig.~\ref{fig:MRplot_SpinPol}).

\section{EOS construction}

We start from the general EOS extension of \cite{Tews:2018kmu,Tews:2018iwm}. This family of EOSs is constrained at nuclear densities by microscopic calculations using local chiral effective field theory (EFT)
interactions and precise Quantum Monte Carlo (QMC) methods, see \cite{Gezerlis:2013ipa,Gezerlis:2014zia,Tews:2015ufa}, and \citep{Lynn:2015jua} for details. 
Chiral EFT provides a systematic theory for nuclear forces, based on the symmetries of quantum chromodynamics, in terms of nucleon and pion degrees of freedom~\citep{Epelbaum2009,Machleidt:2011zz,Hamm13RMP}. 
It explicitly includes long-range pion-exchange interactions, and parameterizes short-range interactions by a general operator basis whose low-energy couplings are fit to nucleon-nucleon ($NN$) scattering data as well as few-body systems. 
Chiral EFT naturally provides three-nucleon (3N) interactions which have been found to be extremely important for calculations of nuclear matter, while four-body interactions have been found to be very small~\citep{Kruger:2013kua,Tews:2012fj}. 
Due to their systematic organization,
chiral interactions can be systematically improved and enable theoretical uncertainty estimates. 
Being a momentum expansion, chiral EFT is limited to low momenta as explored in atomic nuclei, but it allows to 
constrain the EOS at nuclear densities~\citep{Tews:2012fj,Hebeler:2015hla,Gandolfi:2019zpj}.
In addition, QMC methods are among the most precise methods to solve the many-body problem~\citep{Carlson:2014vla}.
Using chiral interactions as input, QMC methods have been used to calculate nuclei and neutron matter with great success~\citep{Lynn:2015jua,Lynn:2019rdt}, which shows
that microscopic calculations can connect the physics of nuclei with the astrophysics of neutron stars.

\begin{figure}[t]
\centering
\includegraphics[width=0.45\columnwidth]{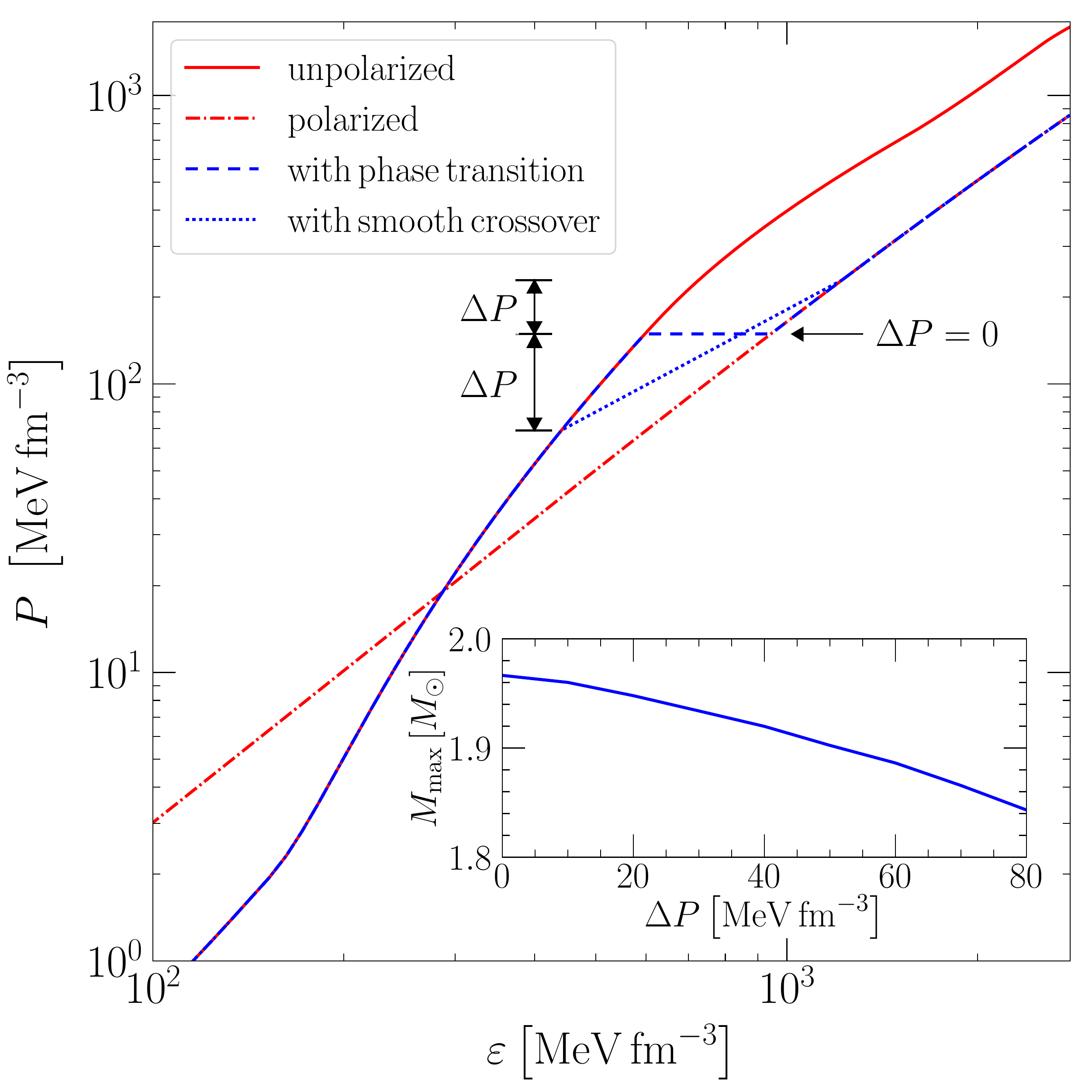}
\caption{\label{fig:Sketch}
Example EOS for unpolarized neutron-star matter (red solid line) and SPM (red dashed-dotted line), and the EOS that results
from a strong first-order phase transition between the two phases in the Maxwell construction (blue dashed line with $\Delta P=0$), as well as an EOS that results when smearing out the phase transition as in a Gibbs construction (blue dotted lines with finite $\Delta P$). Inset:
$M_{\rm{max}}$ for one representative EOS with phase transition to SPM as a function of $\Delta P$.}
\end{figure} 

For the results in this paper, we have 
used the auxiliary-field diffusion Monte Carlo (AFDMC) method~\citep{Schmidt:1999}. 
Together with local chiral EFT interactions, this approach can be applied to neutron matter up to densities around $2 n_{\rm{sat}}$~\citep{Tews:2018iwm}. 
In \cite{Tews:2018kmu,Tews:2018iwm}, we demonstrated how to obtain the neutron-star EOS from these calculations. To extend these calculations to higher densities explored in the neutron-star core, we have used an extension in the speed of sound, $c_S$, which allows to model the most general family of EOSs consistent with our nuclear-density results (see also \cite{Greif:2018njt}). 
We stress that this general extension scheme is independent of a particular choice of degrees of freedom at higher densities. Instead, it explores all allowed density dependencies that are consistent with microscopic calculations at nuclear density, that are causal, and that lead to a stable neutron star, i.e., a star with monotonously growing pressure. As a consequence, this extension scheme covers all possible EOS models, e.g., models with phase transitions, sudden stiffening of the EOS etc.

Using chiral EFT constraints up to $n_{\rm{sat}}$ and the $c_S$ extension at higher densities, we sample tens of thousands of EOSs within the allowed EOS range. For this ensemble of EOSs, we find 
$8.4\km \leqslant R_{1.4}\leqslant 15.2 \km$ for the radius of a typical $1.4 \, M_{\odot}$ neutron star, $R_{1.4}$, and $M_{\rm{max}} \leqslant 4.0 \, M_{\odot}$. 
This upper limit results from the stiffest nuclear EOS consistent with local 
chiral EFT constraints at nuclear densities and the stiffest possible causal EOS at higher densities, and reduces to $2.9 \, M_{\odot}$ if nuclear-physics input is considered up to $2 n_{\rm{sat}}$, as discussed before. 
We show the most general EOS band consistent with nuclear-physics constraints up to $n_{\rm{sat}}$ as gray bands in Fig.~\ref{fig:MRplot_SpinPol}.

\begin{figure*}[t]
\centering
\includegraphics[width=0.925\textwidth]{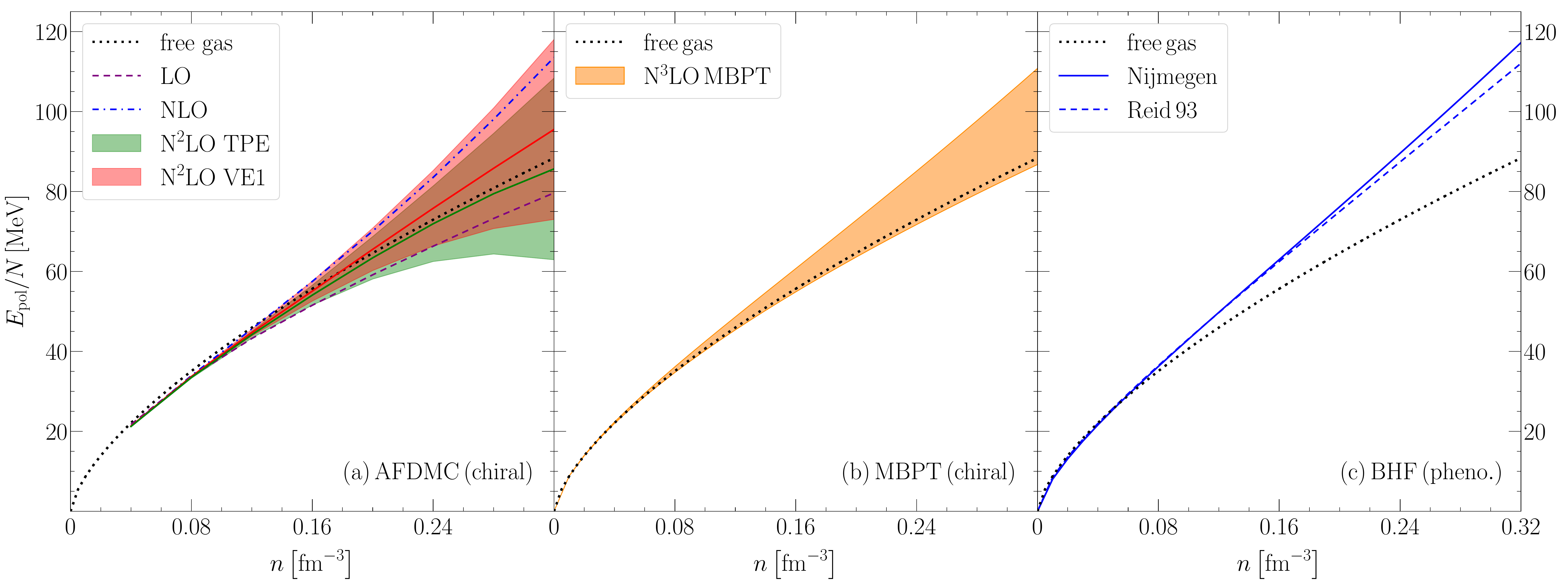}
\caption{\label{fig:EOSplots}
Energy per particle for SPM as a function of density obtained from the three calculations discussed in the text. For comparison, the free spin-polarized gas is shown.
For the AFDMC calculations, we give results for two 3N-force parameterizations at N$^2$LO (TPE and VE1), with the centroid as solid lines and uncertainty bands following \cite{Epelbaum:2014efa}. In the middle panel, the band is obtained by exploring different chiral interactions as well as cutoff and 3N coupling variations~\citep{Kruger:2014caa}.}
\end{figure*}

For each of the EOSs within the EOS band, we construct a new EOS that includes a phase transition to SPM. We sketch our construction in Fig.~\ref{fig:Sketch}. To obtain the EOS for SPM, we use three different calculations: AFDMC calculations using local chiral interactions to next-to-next-to-leading order (N$^2$LO), many-body perturbation theory (MBPT) calculations from \cite{Kruger:2014caa} based on chiral EFT interactions to next-to-next-to-next-to-leading order (N$^3$LO), as well as Brueckner-Hartree-Fock (BHF) calculations~\citep{Vidana:2001ei} based on the phenomenological Reid 93 and Nijmegen~\citep{Stoks:1994wp} potentials. We show the results for SPM for these different calculations in Fig.~\ref{fig:EOSplots}. 

\section{Results}

Using AFDMC, we performed calculations of SPM at leading order (LO), next-to-leading order (NLO), and for two different Hamiltonians at N$^2$LO, which differ by the choice of the 3N parameterization, see \cite{Lynn:2015jua} for details. 
For each Hamiltonian, we estimate the uncertainties from the order-by-order convergence~\citep{Epelbaum:2014efa}, giving the bands shown in Fig.~\ref{fig:EOSplots}~(a). 
Our AFDMC results are in good agreement with the findings of \cite{Riz:2018xmk}, which used the same method and interactions, but did not estimate the EFT uncertainties.
In the following, we will only use the more conservative $V_{E\mathbbm{1}}$ results
(with higher energies for SPM, see the red band in Fig.~\ref{fig:EOSplots}), but note that both 3N parameterizations overlap well. For the $V_{E\mathbbm{1}}$ parameterization, we give the EOS up to $2n_{\rm sat}$ in Table~\ref{tab:EOS}.

\begin{table}[t]
\centering
\label{tab:par}
\begin{ruledtabular}
\begin{tabular}{crrr}
$n\, [\fmiq]$ & $E/N$ [MeV] & $\epsilon\, [\mev \fmiq]$   & $P\, [\mev \fmiq]$ \\
\hline
0.04 & 21.31(7)   & 38.41(1)   & 0.55(1)\\
0.08 & 33.72(11)  & 77.81(1)   & 1.83(5)\\
0.12 & 44.69(83)  & 118.0(1)   & 3.8(4)\\
0.16 & 55.1(2.5)  & 159.0(4)   & 6.5(1.4)\\
0.20 & 65.5(5.3)  & 200.8(1.1) & 10.1(3.4)\\
0.24 & 75.8(9.4)  & 243.4(2.3) & 14.5(6.9)\\
0.28 & 85.8(15.1) & 286.8(4.2) & 19.9(12.4)\\
0.32 & 95.5(22.5) & 331.0(7.1) & 26.3(20.3)\\
\hline
\end{tabular}
\end{ruledtabular}
\caption{\label{tab:EOS}
EOS of SPM from the AFDMC calculation with the N$^2$LO $V_{E\mathbbm{1}}$ parameterization of Fig.~\ref{fig:EOSplots}~(a) up to 2$n_{\rm sat}$.}
\end{table}

As indicated by the uncertainty bands in Fig.~\ref{fig:EOSplots}, chiral interactions become less reliable with increasing density. This is especially true for local interactions employed in AFDMC, because they suffer from sizable local
regulator artifacts~\citep{Dyhdalo:2016ygz,Tews:2015ufa,Huth:2017wzw}. To estimate their impact, we also explore MBPT calculations with nonlocal chiral interactions~\citep{Kruger:2014caa}, which do not suffer from these additional regulator artifacts. 
We show the total MBPT uncertainty band from \cite{Kruger:2014caa}, which was obtained by studying several chiral N$^3$LO interactions as well as variations of the cutoff and the 3N couplings. 
While the uncertainty band is not based on a systematic order-by-order study, the described uncertainty estimation is very reasonable at N$^3$LO, and consistent with the AFDMC calculation, but considerably smaller at larger densities. 
Finally, we compare to results from BHF calculations based on the Nijmegen and Reid93 phenomenological interactions~\citep{Vidana:2001ei}.
These calculations do not provide uncertainties but describe NN scattering data with high precision and are in good agreement with the chiral EFT results. 

To extrapolate the chiral EFT results to higher densities, we fit the simple functional form~\citep{Gandolfi:2009fj}
\begin{equation}
\frac{E_{\rm pol}}{N}(n)= a \cdot \left(\frac{n}{n_{\rm{sat}}} \right)^{\alpha} + b \cdot \left(\frac{n}{n_{\rm{sat}}} \right)^{\beta}
\end{equation}
to the results. In particular, in the case of AFDMC, we fit this functional to the result of our calculation as well as to the upper and lower bounds of the uncertainty bands up to $n_{\rm sat}$. For MBPT, we fit the functional to the bounds of the uncertainty band.
In the case of the AFDMC results, we test the quality of the extrapolation by comparing it to the data points between $n_{\rm sat}$ and $2n_{\rm sat}$ and find that the fit provides a reliable extrapolation to these higher densities. 
For the BHF results, we fit this functional to the individual results over the whole density range. Finally, in case the EOS for spin-polarized matter becomes acausal, we replace it by a causal EOS with $c_s=c$.

As shown in Fig.~\ref{fig:Sketch} for given neutron-star and SPM EOSs, unpolarized matter is energetically favorable at low energy densities but becomes less favorable than spin-polarized matter at higher densities for sufficiently stiff EOSs. 
The reason is that interactions in SPM tend to be weak and results are close to the free Fermi gas~\citep{Kruger:2014caa}, while interactions in unpolarized matter are much stronger and become increasingly repulsive.
We then identify the phase transition between unpolarized matter and SPM by a Maxwell construction, i.e., by matching pressure and Gibbs energy or chemical potential, and construct a new EOS using the unpolarized EOS below and the polarized EOS above the phase transition (blue dashed line in Fig.~\ref{fig:Sketch}). 

We emphasize that this neglects corrections to the spin-polarized matter EOS from protons (or other particle species), which are expected at the level of $\approx 10\,\%$ given typical proton fractions.

To explore the sensitivity to the exact construction of the phase transition, we have explored additional EOSs where the phase transition is smeared out, similar to a Gibbs construction. Such a transition would appear due to the formation of a mixed phase if protons and electrons were included~\citep{Glendenning:1992vb,Heiselberg:1992dx}.  Instead of enforcing $P_{\rm tr}^{\rm polarized}=P_{\rm tr}^{\rm unpolarized}=P_{\rm pt}$, and connecting the EOSs by a segment with $c_s=0$, we construct EOSs with the unpolarized phase up to $P_{\rm pt}-\Delta P$ and the polarized phase after $P_{\rm pt}+\Delta P$. 
These EOS segments are then connected by a smooth interpolation. 
The resulting $M_{\rm max}$ as a function of 
$\Delta P$ is shown in the inset of Fig.~\ref{fig:Sketch} for the given EOS. 
Smearing out the phase transition lowers $M_{\rm max}$ because the EOS gets softened earlier. 
The maximum $M_{\text{max}}$ is therefore found for a strong first-order phase transition resulting from the Maxwell construction. 
This is in good agreement with similar findings for phase transitions to quark matter of \cite{Bhattacharyya:2009fg} and \cite{Wu:2018zoe} and holds for all EOSs in our sample. 
Therefore, our conclusions are robust with respect to the properties of the transition.

We have repeated this construction for all unpolarized neutron-star EOSs in our original band and the different EOSs for SPM: the AFDMC result, its upper and lower bounds, the upper and lower MBPT bounds, as well as the two phenomenological EOSs. This leads
to the results of Fig.~\ref{fig:MRplot_SpinPol}, where the hatched areas are given by the uncertainty bands of the SPM calculations in the corresponding panels of  Fig.~\ref{fig:EOSplots} (or by the two different Hamiltonians in the BHF case).
For the AFDMC calculations, we find $1.75\,  M_{\odot} \leqslant  M_{\rm{max}} \leqslant 2.93 \, M_{\odot}$, with the centroid being at $M_{\rm{max}}=2.29 \, M_{\odot}$. For the MBPT calculations, we find $1.84 M_{\odot} \, \leqslant  M_{\rm{max}} \leqslant 2.59 \, M_{\odot}$, and for the BHF calculations, $ M_{\text{max}} \leqslant 2.61 \, M_{\odot}$ or $ M_{\text{max}} \leqslant 2.44 \, M_{\odot}$. Except in the stiffest possible case, for the upper AFDMC bound, these findings are in very good agreement with each other. However, for this stiffest case the uncertainty is increased by regulator artifacts and most likely overestimated. We find that the predicted $M_{\text{max}}$ is in very good agreement with inferences from the EM counterpart of GW170817 of \cite{Margalit:2017}, \cite{Shibata:2017xdx}, and \cite{Rezzolla:2017aly}, which are shown in Fig.~\ref{fig:MRplot_SpinPol} as horizontal lines or bands. 

Because a phase transition to SPM softens the EOS drastically, we find that it is unlikely that a neutron-star with a spin-polarized core exists in nature. Typically, we find the mass of the SPM domain to be $\leqslant 0.005 \, M_{\odot}$, largely a result of numerical discretization artifacts.
Only the stiffest possible spin-polarized EOSs can stabilize any star with spin-polarized matter in their core. 
In this case, we find that the mass of spin-polarized domain in the neutron-star core is $\leqslant 0.19 \, M_{\odot}$. However, most of the resulting EOSs lead to neutron stars with $R_{1.4} \geqslant 13.6$\,km and are therefore ruled out by GW170817~\citep{Abbott:2017,Abbott:2018wiz,Annala:2017llu,Most:2018hfd,Tews:2018iwm}.
If we were to exclude these EOSs from consideration, 
we find that the mass of spin-polarized matter in the core is $\leqslant 0.02 \, M_{\odot}$. 

We have ignored the effects of magnetic fields, which could impact the EOS of spin-polarized matter if there is a net magnetization. 
We have also ignored a gradual polarization of neutron matter which, however, softens the EOS sooner, leading to a lower $M_{\rm max}$. Hence, the case we investigated presents an upper limit on $M_{\rm max}$ due to a transition to SPM in the core. 
Finally, a similar effect on the EOS might occur due to a phase transition to deconfined quark matter; see, e.g., category A in \cite{Alford:2013aca}. However, since quark-matter properties cannot be predicted from first principles in the density range of interest, in this case no strong constraint on $M_{\rm max}$ can be obtained.

\section{Summary}
We have investigated the impact of a phase transition from neutron-star matter to SPM and found that such a phase transition limits the pressure in the neutron-star core. 
Combining information from AFDMC calculations, as well as previous MBPT~\citep{Kruger:2014caa} and BHF~\citep{Vidana:2001ei} calculations limits $M_{\rm max}$ of neutron stars to lie below 2.6$-$$2.9 \, M_{\odot}$, depending on the microscopic nuclear forces used, while significantly larger $M_{\rm max}$ of 3$-$$4\,M_{\odot}$ could be reached without these constraints. These limits can be improved if the uncertainty in SPM calculations is reduced. 
The lower $M_{\rm max}$, and in particular the result for the AFDMC calculations without uncertainty estimates, $M_{\rm{max}} = 2.29 \, M_{\odot}$,
are in very good agreement with recent constraints from the EM counterpart of GW170817 from \cite{Margalit:2017}, \cite{Shibata:2017xdx}, and  \cite{Rezzolla:2017aly}. Finally, we find that stars with a large spin-polarized domain in their core are ruled out by the radius constraint from GW170817.

\acknowledgments

We are grateful to C.J.~Pethick for helpful discussions and J.~Carlson and C.~Wellenhofer for comments on the manuscript. This work was supported by the US Department of Energy, Office of Science, Office of Nuclear Physics, under Contract DE-AC52-06NA25396, the Los Alamos National Laboratory (LANL) LDRD program, the NUCLEI SciDAC  program, and the Deutsche Forschungsgemeinschaft (DFG, German Research Foundation) -- Projektnummer 279384907 -- SFB 1245. Computational resources have been provided by Los Alamos Open Supercomputing via the Institutional Computing (IC) program, by the National Energy Research Scientific Computing Center (NERSC), which is supported by the U.S. Department of Energy, Office of Science, under contract DE-AC02-05CH11231, and by the J\"ulich Supercomputing Center.

\bibliography{bibliography}
\bibliographystyle{aasjournal}

\end{document}